\documentclass[aip,rsi,amsmath,amssymb,reprint,onecolumn]{revtex4-2}

\usepackage{graphicx}
\usepackage{dcolumn}
\usepackage{bm}

\usepackage[utf8]{inputenc}
\usepackage[T1]{fontenc}
\usepackage{mathptmx}
\usepackage{etoolbox}
\usepackage{xcolor}
\usepackage{listings}
\definecolor{codegreen}{rgb}{0,0.6,0}
\definecolor{codegray}{rgb}{0.5,0.5,0.5}
\definecolor{codepurple}{rgb}{0.58,0,0.82}
\definecolor{backcolour}{rgb}{0.95,0.95,0.92}
\lstdefinestyle{mystyle}{
    backgroundcolor=\color{backcolour},   
    commentstyle=\color{codegreen},
    keywordstyle=\color{magenta},
    numberstyle=\tiny\color{codegray},
    stringstyle=\color{codepurple},
    basicstyle=\ttfamily\footnotesize,
    breakatwhitespace=false,         
    breaklines=true,                 
    captionpos=b,                    
    keepspaces=true,                 
    numbers=left,                    
    numbersep=5pt,                  
    showspaces=false,                
    showstringspaces=false,
    showtabs=false,                  
    tabsize=2
}

\makeatletter
\def\@email#1#2{%
 \endgroup
 \patchcmd{\titleblock@produce}
  {\frontmatter@RRAPformat}
  {\frontmatter@RRAPformat{\produce@RRAP{*#1\href{mailto:#2}{#2}}}\frontmatter@RRAPformat}
  {}{}
}%
\makeatother
\begin{document}

\preprint{AIP/123-QED}

\title{Arduino based interferometer stabilizer equipped with a digital lock-in amplifier}

\author{Giuseppe Emanuele Lio}
\affiliation{Istituto Nanoscienze – CNR, NEST-SNS, Piazza San Silvestro 12, 56127 Pisa, Italy}
\affiliation{Dipartimento di Fisica E. Fermi, Università di Pisa, Largo Bruno Pontecorvo 3, 56127, Pisa, Italy}
\author{Saravanan Rajamani}
\affiliation{Istituto Nanoscienze – CNR, NEST-SNS, Piazza San Silvestro 12, 56127 Pisa, Italy}
\author{Alessandro Pitanti}
\affiliation{Dipartimento di Fisica E. Fermi, Università di Pisa, Largo Bruno Pontecorvo 3, 56127, Pisa, Italy}
\affiliation{Istituto Nanoscienze – CNR, NEST-SNS, Piazza San Silvestro 12, 56127 Pisa, Italy}
\author{Stefano Roddaro}
\affiliation{Dipartimento di Fisica E. Fermi, Università di Pisa, Largo Bruno Pontecorvo 3, 56127, Pisa, Italy}
\affiliation{Istituto Nanoscienze – CNR, NEST-SNS, Piazza San Silvestro 12, 56127 Pisa, Italy}
\author{Simone Zanotto}
\affiliation{Istituto Nanoscienze – CNR, NEST-SNS, Piazza San Silvestro 12, 56127 Pisa, Italy}
\email{simone.zanotto@nano.cnr.it}

\date{\today}

\begin{abstract}
Interferometric sensors are ubiquitous in precision metrology, yet their performance is fundamentally limited by environmental noise and thermal drift. To achieve maximum sensitivity, these systems must be actively stabilized at the quadrature point of the interference fringe. Commercial stabilization solutions, typically based on analog lock-in amplifiers or FPGA architectures, are often prohibitively expensive and do not not offer the flexibility needed for custom experimental setups. In this work, we present an open-source, compact and low-cost digital stabilization system built upon the dual-core Arduino Giga R1 microcontroller. The system features a custom analog front-end with programmable gain amplifiers (PGAs) and active signal conditioning, enabling direct integration with standard amplified photodiodes. We implement a firmware-based digital lock-in amplifier running at a 100 kHz sampling rate, which performs real-time demodulation and PID feedback control without the latency bottlenecks of PC-based loops. Experimental characterization demonstrates that the system effectively suppresses long-term thermal drift and actively rejects external perturbations. The resulting device provides a standalone, Arduino-based alternative for laser frequency stabilization and interferometric control in educational and research laboratories.
\end{abstract}

\maketitle

\section{\label{sec:intro}Introduction}
Interferometry stands as the premier method for precision displacement measurement across diverse fields of science and technology, ranging from the nanometer scale such as in metrological atomic force microscopy \cite{rugar1989improved, zhang2017quantitative, muhlberger2021high, de2022simultaneous, lee2025roadmap, liang2025modulated} to large-scale such as gravitational wave detection \cite{rugar2004single, belfi2014interferometric, tsaturyan2017ultracoherent} and radiation pressure on light-sails \cite{santi2023swarm, michaeli2025direct}. The ability to resolve displacements down to picometer precision with nanometer accuracy has established interferometric techniques as the gold standard in modern metrology \cite{bobroff1993recent, canuto2007digital, vsmid2008precision, niwa2009long, cordero2009interferometry, schuldt2012picometre, yang2018review}.
However, the same extreme sensitivity that allows for such precision also renders interferometers highly susceptible to environmental instabilities, a phenomenon often described as the ``zero-drift'' problem \cite{von2025highly}. This drift encompasses all slow, unintentional variations in the interferometric signal primarily caused by thermal expansion, mechanical creep, or laser wavelength instability \cite{wehner1997scanning, vu2016accurate, hu2019displacement, von2025highly}. In basic, unmodulated setups, these variations directly mimic low-frequency target displacements. Conversely, in systems utilizing frequency demodulation or spectral analysis, while the drift can be spectrally distinguished from the dynamic signal, it remains highly detrimental; it shifts the system operating point, thereby distorting the transfer function and inducing non-linear responses \cite{chen2018phase, zhang2023high, liu2025high}.
For a two-beam interferometer, such as the Michelson configuration, this drift is particularly detrimental. To achieve maximum detection sensitivity and linearity, the system must be maintained at the ``quadrature'' point (the center of the interference fringe slope). Uncompensated thermal drift inevitably pushes the system towards the interference extrema (minima or maxima), where the sensitivity to small displacements vanishes and the signal response becomes non-linear. While various solutions exist to mitigate these effects ranging from passive thermal isolation to cryogenic operation, these approaches are often bulky, expensive, or limited to specific environments \cite{lipsett1966laser, krishnamachari2006active, aso2004stabilization, zhang2005optical, cho2009stabilization, wang2024wavelength}. Active feedback control remains the most robust solution for room-temperature applications, yet commercial laser locking controllers can be prohibitively expensive and lack the flexibility required for custom experimental setups \cite{flugge2009interferometry, gerberding2017laser, eichholz2015heterodyne, pitanti2015strong, ruksasakchai2022microcontroller, pollastrone2023fully, harvie23}. Stabilized interferometry can also be essential when analyzing optomechanical systems, where micro and nano-metric resonators are natively coupled with photonic modes \cite{pitanti2015strong, greiffenhagen2019discussion, baldacci2016thermal, shao2022electrical, wilson2015measurement, koh2025decoherence, pitanti2025ghz}.
In this work, we aim at filling the gap left by the dominance of high-performance commercial hardware for interferometer stabilization. For many applications, indeed, the specifications of such systems are often exceedingly high; moreover, the closed nature of their software and hardware makes them hardly - if not impossible - to adapt to new configurations. We hence present a standalone digital stabilization unit based on the Arduino Giga R1, a dual-core microcontroller that combines the ease of high-level programming with the deterministic timing required for analog signal processing. Unlike simple DC-locking schemes, our system implements a digital lock-in amplifier algorithm coupled with a PID (proportional-integral-derivative) feedback loop. By introducing a 1 kHz dithering perturbation to the mirror position and sampling the response at 100 kHz, we extract a drift-free error signal that is robust against intensity fluctuations. We describe the complete hardware design — including a custom signal conditioning shield with active filtering and Programmable Gain Amplifiers (PGAs) - and demonstrate the system ability to lock a Michelson interferometer to its quadrature point for extended periods, rejecting both slow thermal drift and fast mechanical vibrations.
\section{\label{sec:theory} Operation principles}

\begin{figure}
    \centering
    \includegraphics[width=9 cm]{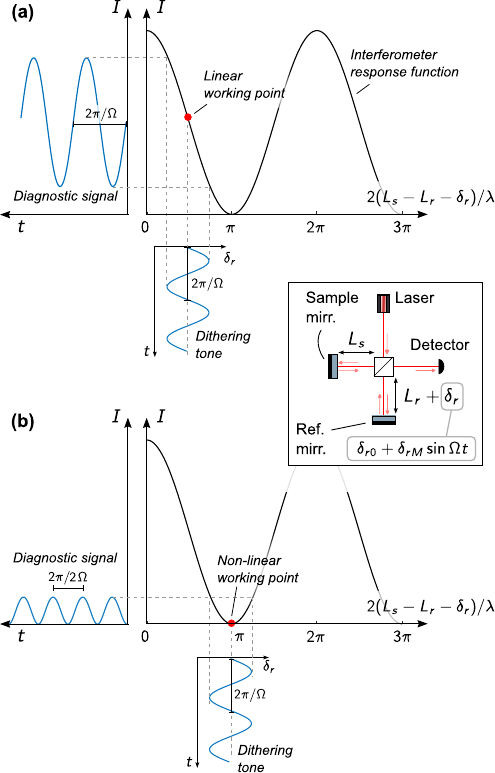}
    \caption{Principles of interferometer stabilization at the linear working point. Since the setup-intrinsic optical path difference $L_s-L_r$ is in general different from the ideal condition $2(L_s-L_r)/\lambda = (2m+1)\pi/2$, an intentional, quasi-static additional offset $\delta_{r,0}$ is superimposed to drive the reference mirror to the desired position. To determine whether the optical path difference actually matches the ideal condition, a dithering tone $\delta_{r,M} \sin \Omega t$ is further superimposed. If $2(L_s-L_r-\delta_{r,0})/\lambda = (2m+1)\pi/2$, the interferometer is at the linear working point, and the dithering at $\Omega$ results in a diagnostic signal with a pulsation $\Omega$ [panel (a)]. If the interferometer is far from the linear working point [with $2(L_s-L_r-\delta_{r,0})/\lambda = m\pi$ as a worst-case scenario, panel (b)], the dithering at $\Omega$ results in a diagnostic signal with a pulsation $2\Omega$. In general, the diagnostic signal is a superposition of $\Omega$ and $2\Omega$, and a Fourier analysis allows to determine the actual value of $\phi = 2(L_s-L_r-\delta_{r,0})/\lambda$.}
    \label{fig:InterfScheme}
\end{figure}

The core of a Michelson interferometer operation (Figure \ref{fig:InterfScheme}) is based on the following response function:
\begin{eqnarray}
\label{eq:interf1}
    I & = & \frac{I_0}{2} \left( 1 + \cos\left( \frac{2(L_s + \delta_s) - 2(L_r + \delta_r)}{\lambda}  \right) \right) \\ \nonumber
              & = &\frac{I_0}{2} \left( 1 + \cos\left( \frac{2\Delta L}{\lambda} -  \frac{2\delta_r}{\lambda} + \frac{2\delta_s}{\lambda}     \right) \right) 
\end{eqnarray}
where $I$ is the light intensity at the detector, $I_0$ is the laser intensity, $\lambda$ is the laser wavelength, $L_{s,r}$ are respectively the lengths of sample and reference arms, $\Delta L = L_s-L_r$ is the undesired beam path difference (i.e.~the quantity that is subject to mounting imperfections and to uncontrollable drifts), $\delta_s$ is the length change of the the sample arm (i.e.~the quantity to be measured), and $\delta_r$ is a controllable variation of the reference arm length \footnote{ https://www.zhinst.com/sites/default/files/zi\_mfli\_appnote\_interferometry.pdf }. Analyzing such response function highlights two issues: first, it is in general nonlinear with respect to $\delta_s$; second, it depends upon $\Delta L$, which is a time-dependent quantity because of drifts. The purpose of interferometer stabilization is thus to act upon $\delta_r$ in order to compensate for the drifts in $\Delta L$, as well as to lock the interferometer at a linear working point. This can be accomplished assuming that the sample motion is smaller than the wavelength ($\delta_s \ll \lambda$), and that the reference arm is tuned such as $2\Delta L/\lambda-2\delta_r/\lambda = (2m+1)\pi/2$, where $m$ is an integer. In this case, Eq.~(\ref{eq:interf1}) reduces to
\begin{equation}
    I = I_0 \left( \frac{1}{2} \pm \frac{\delta_s}{\lambda} \right)
    \label{eq:LinearResponse}
\end{equation}
where + (-) applies for odd (even) values of $m$.

To determine the appropriate value of $\delta_r$, let's assume that the sample arm is at zero displacement ($\delta_s = 0$). Let's also assume that $\delta_r$ is driven as a superposition of a quasi-static displacement $\delta_{r,0}$ and a time-harmonic displacement modulation of amplitude $\delta_{r,M}$ and pulsation $\Omega$ (dithering). The detector will hence receive an intensity given by
\begin{eqnarray}
    I(t) & = & \frac{I_0}{2} \left( 1 + \cos \left(  \phi - \frac{2\delta_{r,M}}{\lambda} \sin (\Omega t + \psi) \right) \right) \nonumber \\
         & = & \frac{I_0}{2} - I_0\, J_1\left(\frac{2\delta_{r,M}}{\lambda}\right)\, \sin \phi\, \sin(\Omega t + \psi)\nonumber \\
         & & \quad \quad + I_0\, J_2\left(\frac{2\delta_{r,M}}{\lambda}\right)\, \cos \phi\, \cos(2\Omega t + \psi)
\label{eq:interf2}
\end{eqnarray}
where $\phi = 2(\Delta L - \delta_{r,0} )/\lambda $, and $J_{1,2}$ are the first and second Bessel functions of the first kind. The phase $\psi$ originates from the signal propagation through the analog path.  A graphical representation of Eq.~\ref{eq:interf2} is given in Fig.~\ref{fig:InterfScheme}. If $\phi = (2m+1)\pi/2$ (we report the case $m=0$ in Fig.~\ref{fig:InterfScheme}a), the sinusoidal dithering at $\Omega$ results in a sinusoidal signal at the same $\Omega$. If $\phi = m\pi$ (we report the case $m=1$ in Fig.~\ref{fig:InterfScheme}b), the dithering at $\Omega$ results in a doubled frequency signal. In general, the signal $I(t)$ is a superposition of signals at $\Omega$ and at $2\Omega$, depending on the value of $\phi$; we will leverage on this property to determine the actual value of $\phi$ and to drive the reference mirror to a position such that $\phi$ assumes the form $(2m+1)\pi/2$ for a certain integer $m$. 
To operate the stabilization, the light intensity is converted into a photovoltage, that is subsequently filtered and amplified to yield the time-dependent voltage available at the \texttt{A1} analog-to-digital converter input (details below). As the signal chain introduces frequency-dependent amplitude scaling factors and phase delays, the digitized voltage is
\begin{eqnarray}
    V_{\mathtt{A1}}(t) & = & -k_{\Omega}\, I_0\, J_1\left(\frac{2\delta_{r,M}}{\lambda}\right)\, \sin \phi\, \sin(\Omega t + \psi_{\Omega})\nonumber \\
         & & + k_{2\Omega}\, I_0\, J_2\left(\frac{2\delta_{r,M}}{\lambda}\right)\, \cos \phi\, \cos(2\Omega t + \psi_{2\Omega})
         \label{eq:VA1}
\end{eqnarray}
where $k_{\Omega, 2\Omega}$ and $\psi_{\Omega, 2\Omega}$ are fixed, hardware-dependent constants. 
To determine $\phi$, we resorted to the lock-in detection scheme applied to $V_{\mathtt{A1}}(t)$, which yields first- and second-harmonic quadratures  
\begin{eqnarray}
    X_1 & = & -k_{\Omega}\, I_0\, J_1^*\, \sin \phi\, \sin \psi_{\Omega}\nonumber \\
    Y_1 & = & -k_{\Omega}\, I_0\, J_1^*\, \sin \phi\, \cos \psi_{\Omega}\nonumber \\
    X_2 & = & k_{2 \Omega}\, I_0\, J_2^*\, \cos \phi\, \cos \psi_{2 \Omega}\nonumber \\
    Y_2 & = & k_{2 \Omega}\, I_0\, J_2^*\, \cos \phi\, \sin \psi_{2 \Omega}
\label{eq:XY}
\end{eqnarray}
where $J_{1,2}^* = J_{1,2}(2\delta_{r,M}/\lambda)$. The lock-in amplitudes $R_{1,2} = \sqrt{X_{1,2}^2+Y_{1,2}^2}$ can then be defined, and result in
\begin{eqnarray}
    R_1 & = & k_{\Omega}\, I_0\, J_1^*\, |\sin \phi\,| \nonumber \\
    R_2 & = & k_{2 \Omega}\, I_0\, J_2^*\, |\cos \phi\,| .
\label{eq:R}
\end{eqnarray}
To determine the unknown prefactors $k_{\Omega}\, I_0\, J_1^*$ and $k_{2\Omega}\, I_0\, J_2^*$ we devised a calibration protocol. The procedure simply consists of sweeping the quasi-static displacement $\delta_{r,0}$, which implies a sweep of $\phi$; since the range is by design wider than $2\pi$, the maximum measured $R_{1,2}$ values directly determine the prefactors. The final step consists of taking a four-quadrant inverse tangent, using also the information contained in $X_{1,2}$:
\begin{equation}
    \tilde{\phi} = \mathrm{atan2}\left( -\frac{R_1}{\max_{\delta_{r,0}} R_1}\,\mathrm{sign} X_1,\ \frac{R_2}{\max_{\delta_{r,0}} R_2}\, \mathrm{sign} X_2  \right).
    \label{eq:phitilde}
\end{equation}
We indicated the result of this operation with a tilde since a piece of information is lost in this process. Indeed, one has $\tilde{\phi} = \pm \phi$ or $\tilde{\phi} = \pm \phi+\pi$, depending on the unknown signs of $\sin \psi_{\Omega}$ and $\cos \psi_{2 \Omega}$. However, this has no impact on the final goal, since the linear working point occurs when $\phi$ is of the form $(2m+1)\pi/2$ for any $m$ integer. In other words, having experimental access to $\tilde{\phi}$, one can drive the reference mirror by a feedback mechanism to set $\tilde{\phi} = \pi/2$, which implies either $\phi = \pi/2$ or $\phi = -\pi/2$, and ultimately a linear interferometer response as in Eq.~\ref{eq:LinearResponse}.

The procedure described here could in principle be simplified by noticing that the linear operating condition $\phi = (2m+1)\pi/2$ is satisfied when $R_2 = 0$. However, it is advisable not to rely on the sole $R_2$ information to drive a feedback mechanism. At first, $R_2$ is affected by laser intensity fluctuations (either originating from the source or from any element in the optical path), leading to possibly inappropriate feedback signals. Moreover, using the sole $R_2$ entails a non-smooth dependence stemming from the modulus in Eq.~\ref{eq:R}. Instead, using Eq.~\ref{eq:phitilde} leads to a continuous and smooth functional dependence, that is by far more appropriate as an input to the subsequent PID feedback stage.

The PID feedback is implemented with the following expression
\begin{equation}
    f_{\mathrm{PID},i} = 0.5+ K_P\, e_i + K_I\, \Delta t\, \sum_{j<i} e_j  + K_D \frac{e_i-e_{i-1}}{\Delta t} 
    \label{eq:pid}
\end{equation}
where  $f_{\mathrm{PID},i}$ is the feedback at iteration $i$, intended in the [0,1] range; $e_i = \tilde{\phi}_i-\phi_{\mathrm{targ}}$ is the error at iteration $i$; $\phi_{\mathrm{targ}}$ is the target phase (usually $90^{\circ}$); $\Delta t$ is the time interval between successive PID iterations (usually 50 ms, which is several times the lock-in time constant yet small enough to have a responsive feedback). $K_P$, $K_I$, $K_D$ are the PID constants, whose tuning is discussed later. The integral sum $\sum_{j<i} e_j$ is reset to 0 every time the feedback $f_{\mathrm{PID},i}$ exceeds bounds given by $f_{\mathrm{PID},i} < a$ or $f_{\mathrm{PID},i} > 1-a$, where $a$ is the amplitude of the dithering tone, also intended to be in the [0,1] range. 

\section{\label{sec:expset}Experimental Setup} 

\subsection{\label{sec:expset_pcb}Electronic Design and Signal Conditioning}
\begin{figure*}[h]
    \centering
    \includegraphics[width=1\linewidth]{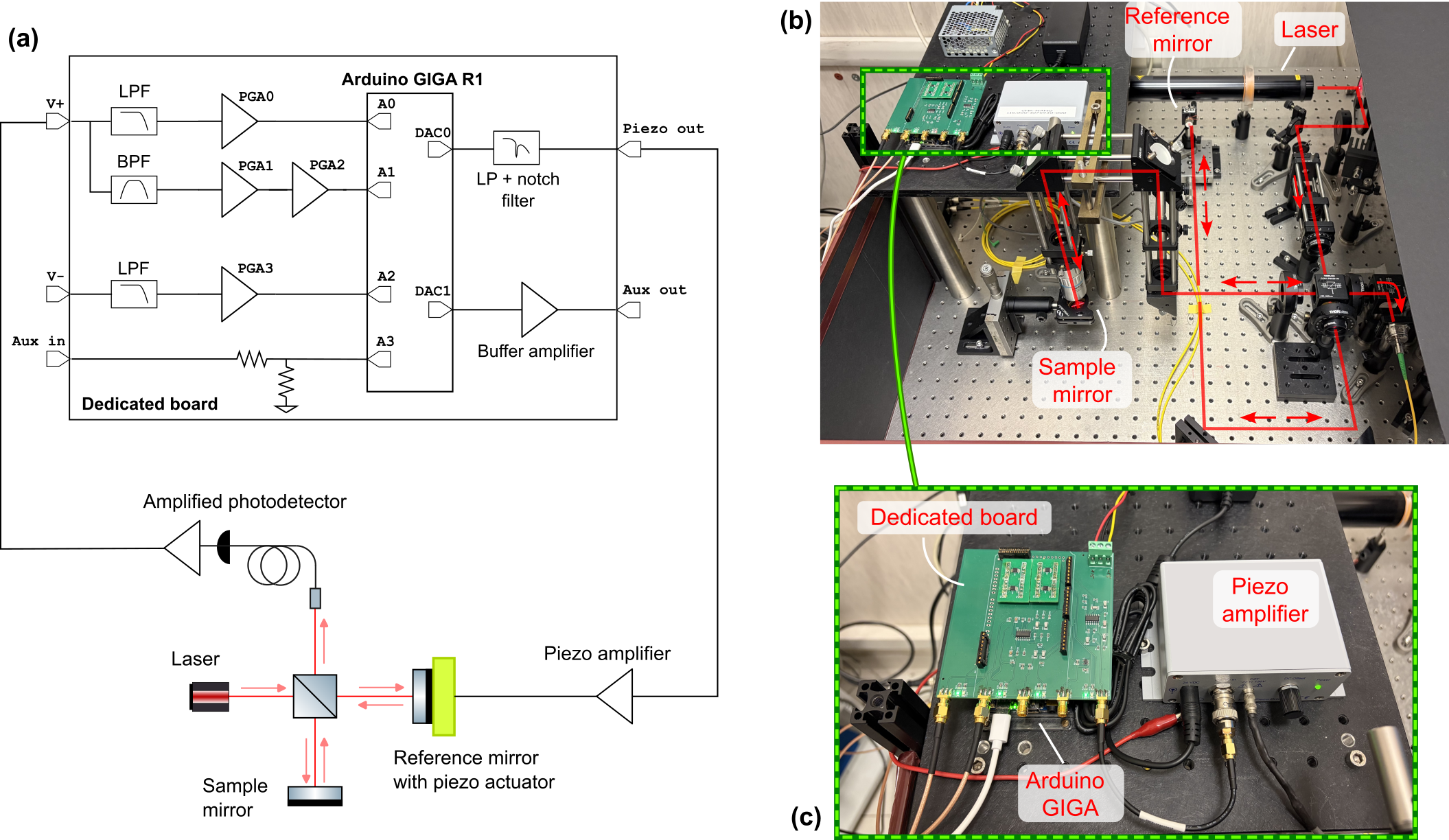}
    \caption{Experimental implementation of the active stabilization system. (a) detailed block diagram of the electronic control loop and of the optical interferometer. The optical signal detected by the photodetector is processed by a custom analog front-end before digitization. The signal path includes analog filtering stages (low-pass and band-pass) to separate DC and AC components, followed by Programmable Gain Amplifiers (\texttt{PGA}) that optimize the dynamic range for the analog inputs (\texttt{A0-A1}) of Arduino Giga R1 microcontroller. The microcontroller digital-to-analog port (\texttt{DAC0}) is filtered and externally amplified to drive the piezoelectric mirror actuator. An additional photovoltage input (\texttt{V-}) an auxiliary signal input (\texttt{Aux in}) and an auxiliary output (\texttt{Aux out}) have been included in the board for compatibility with advanced interferometric techniques, to provide back-up operation with an external lock-in, and for debug purposes (see text). (b) Photograph of the optical bench setup, showing the interferometer and the stabilization hardware. (c) Detail of the custom-made electronic shield (green PCB) housing the conditioning filters and PGAs, mounted directly onto the microcontroller.}
    \label{fig2}
\end{figure*}

The core of the stabilization loop is a hybrid analog-digital control unit built around the Arduino Giga R1 microcontroller. We selected this platform for its high-performance dual-core architecture and its 12-bit analog input/output resolution, which is sufficient for high-fidelity locking. To ensure robust operation across different optical intensity ranges, we developed a custom electronic interface board (shown in Figure \ref{fig2}b-c) that bridges the optical detector with the digital logic. 

The circuit block schematic is represented in Figure \ref{fig2}a. Starting from the detection branch, the signal from the photodetector (Thorlabs PDB450A) undergoes a multi-stage conditioning process before reaching the microcontroller’s Analog-to-Digital Converters (ADCs). First, the analog input (port \texttt{V+}) is split into parallel paths to isolate the relevant spectral components. The DC path extracts the average optical power, allowing the system to monitor beam alignment and slow intensity fluctuations independently of the interferometric signal. The low-pass filter is a Sallen-Key second-order with 1 V/V DC gain and -3dB frequency at 20 Hz (see Appendix \ref{app:circuits} for design details). The cutoff frequency has been chosen to properly reject the interferometric dithering tone which in our design is centered at  $\Omega/(2\pi) = 1\,\mathrm{kHz}$, while retaining the capability to follow intensity variations on the timescale of a fraction of a second, as usually found while performing laser alignment. The AC path deploys a fourth-order Butterworth filter centered at $1.5\,\mathrm{kHz}$ with a $3\ \mathrm{kHz}$ $-3\ \mathrm{dB}$ passband. The filter has an in-band design gain of 7 V/V, resulting in a gain of 6.5 V/V at both $\Omega$ and $2\Omega$ (see Appendix \ref{app:circuits} for design details). We included in the board a second analog input, \texttt{V-}, that can be connected to the other output of the differential photodetector, for future use with advanced interferometric schemes such as that illustrated in Ref.~\cite{barg2017}. 

After the active analog filters, each signal path is equipped with one or two MAX9939 Programmable Gain Amplifier (PGA). These chips allow the microcontroller to dynamically adjust the signal gain (via SPI communication) to match the full $0-3.3\, \mathrm{V}$ input range of the Arduino ADCs. This gain control, that in further development could be automatized, is vital for maximizing the signal-to-noise ratio without saturation of the digital inputs, effectively adapting the hardware sensitivity to the specific level of input signal and interferometric oscillation amplitude.

Once digitized at the \texttt{A0}-\texttt{A2} ports, the signals are processed by the Arduino Giga R1. The slowly varying signals at \texttt{A0} and \texttt{A2} are sensed for overrange and stored for communication to the graphical user interface (GUI). The rapidly varying signal at \texttt{A1} is elaborated in real time to compute the lock-in variables $X_{1,2}$, $Y_{1,2}$, the phase $\tilde{\phi}$, and the feedback $f_{\mathrm{PID}}$; see Sect.~\ref{sect:firmware} for algorithmic details. Meanwhile, the microcontroller generates the dithering tone with amplitude $a$ which is digitally superimposed to a voltage offset $f$, resulting in a voltage output 
\begin{equation}
V_{\mathtt{DAC0}} = 3.3\times(f + a \sin \Omega t)
\label{eq:piezodrive}
\end{equation}
available at the digital-to-analog port \texttt{DAC0}. Here, $f$ and $a$ are in the $[0,1]$ range. Depending upon the operation, the offset $f$ can be linearly swept in time (calibration operation), set to mid-scale (open-loop configuration), suddenly changed (test of lock-in response time), or set to the PID feedback $f_{\mathrm{PID}}$ (Eq.~\ref{eq:pid}). The signal of Eq.~\ref{eq:piezodrive} is then filtered to remove the quantization noise and the spurious second harmonic content. The filter is a three-stage Antoniou notch followed by a low-pass multiple feedback and an inverting gain block, resulting in a gain of 3.2 V/V at DC, 3.03 V/V at 1 kHz, and 0.026 V/V at 2 KHz (tolerance-averaged design values; see Appendix \ref{app:circuits} for details). The amplified signal is exposed at the \texttt{Piezo out} connector. In our implementation, the signal is further boosted by an external high-voltage amplifier (PI E836) to drive the piezoelectric element glued to the reference mirror (PI PL088.3x multilayer actuator). Considering the amplification factors and the phases introduced by the amplifiers, the piezo mirror is eventually driven by a displacement $\delta_{r,0} + \delta_{r,M} \sin (\Omega t + \psi)$, with $\delta_{r,0} \propto f$ and $\delta_{r,M} \propto a$. Given the hardware characteristics, we have that a full-scale sweep of $f$ in the [0,1] range corresponds to a sweep of $\delta_{r,0} \approx 6\pi \lambda $, and that an amplitude $a = 0.04$ (a typically employed value) gives $\delta_{r,M} \approx 0.3 \lambda $ (see Figures \ref{fig5}-\ref{fig6}).

We have also equipped the board with an auxiliary input/output pair that can be used to bypass the microcontroller-based lock-in, leaving only the PID processing on the Arduino. This auxiliary channel could prove useful if the interferometric signal is too weak or too noisy, making it necessary to rely on a low-noise, high-sensitivity lock-in apparatus.  The auxiliary input \texttt{Aux in} is fed to the \texttt{A3} Arduino ADC through a resistive partitioner to adapt the usual 10V full scale signal to the internal 3.3V logic. The auxiliary output \texttt{Aux out}, fed by \texttt{DAC1}, provides a buffered and amplified square wave output with $5V_{pp}$, dedicated to drive the TTL trigger usually found in standalone lock-in amplifiers. Besides these functions, the auxiliary ports can be used for other applications such as reading and generating test and debug signals, with a rather high level of generality thanks to the absence of frequency filtration.

\subsection{Firmware Architecture and Control Algorithm}
\label{sect:firmware}

The firmware architecture is summarized in Figure \ref{fig3}. After standard function and variable definitions, the  Arduino \texttt{setup()} is called, with code blocks dedicated to hardware initialization and to lookup table creation. Specifically, the microcontroller pre-calculates sine function lookup tables for first and second harmonics and stores it in the RAM as floating point array variables. This avoids computationally expensive trigonometric function calls during the critical execution path.
Afterwards, the Arduino \texttt{loop()} method is called. Here, we implement all the runtime operations, relying on the Arduino global time-base counter for proper synchronization. To maintain high-fidelity signal processing without the latency of a full operating system, the firmware is divided into four primary operational tasks.
The first task (\textbf{T1}) represents the core of the lock-in detection scheme, and is executed as soon as the elapsed time exceeds 10 $\mu$s (100 kHz). In this interval, the system executes three synchronous operations: (i) it updates \texttt{DAC0} with the next point of the modulation sine wave to drive the piezo mirror; (ii) it acquires the instantaneous photodetector voltage via \texttt{A1}; and (iii) it performs real-time demodulation. This is achieved by mixing the input signal with the stored sine and cosine reference tables and subsequently applying a two-stage cascaded Infinite Impulse Response (IIR) low-pass filter. In details, at each call the processor performs the operations listed in Fig.~\ref{fig:LIcode}.
\begin{figure}[h]
\begin{lstlisting}[language=c]
  int sample_index = (micros() % PERIOD) / SAMPLE_INTERVAL_US
  
  float ref1_cos = (float)(sine_wave_table[(sample_index + NUM_SAMPLES / 4) % NUM_SAMPLES]); 
  float ref1_sin = (float)(sine_wave_table[sample_index]);
  float vin = analogRead(analogInput);  // Reading the analog values from A1 

  float X1 = vin * ref1_cos;   // Signal multiplied by X reference
  float Y1 = vin * ref1_sin;   // Signal multiplied by Y reference

  X1_f1 += alpha * (   X1 - X1_f1); // update X, 1st harmonic, 1st filter
  X1_f2 += alpha * (X1_f1 - X1_f2); // update X, 1st harmonic, 2nd filter
  Y1_f1 += alpha * (   Y1 - Y1_f1); // update Y, 1st harmonic, 1st filter
  Y1_f2 += alpha * (Y1_f1 - Y1_f2); // update Y, 1st harmonic, 2nd filter
\end{lstlisting}
\caption{\label{fig:LIcode} Code fragment from the lock-in firmware.}
\end{figure}

Analogous operations are performed using the second harmonic reference. The variables \texttt{X1\_f2}, \texttt{Y1\_f2}, \texttt{X2\_f2}, \texttt{Y2\_f2} represent hence the $X_1$, $Y_1$, $X_2$, $Y_2$ lock-in quantities. Regarding the choice of the filter coefficient, we usually adopted a value $\alpha = 6\times 10^{-4}$, which, at the system sampling rate of 100 kHz, is equivalent to setting the lock-in integration time to $\tau \approx$ 18 ms (corresponding to a -3 dB cutoff frequency of $\approx$10 Hz). This  bandwidth choice effectively suppresses the modulation carrier and high-frequency noise, without hindering proper operation of the subsequent PID task. The effect of different $\alpha$ values is discussed in Appendix \ref{app:characterization}. 

The second task (\textbf{T2}, called as soon as the elapsed time exceeds 51 ms) performs the following operations: (i) it calculates $R_{1,2}$ and the optical phase estimate $\tilde{\phi}$ using Equations \ref{eq:R}-\ref{eq:phitilde}, (ii) it calculates the Proportional-Integral-Derivative (PID) feedback value using Eq.~\ref{eq:pid}. The PID execution occurs only if a specific flag is enabled; in this case, the feedback value $f$ is also used to offset the piezo mirror drive signal, Eq.~\ref{eq:piezodrive}. If the flag is disabled, the piezo mirror offset is set to mid-scale.

The third task (\textbf{T3}, called if a \texttt{monitor\_flag} is on and if the elapsed time has exceeded 500 ms), the microcontroller prints to the serial port a table containing a waveform of the data acquired from \texttt{A1} in the T1 loop; moreover, it plots to the serial the quasi-DC readings (\texttt{A0} and \texttt{A2}).

Last but not the least, another two fundamental processes are the calibration and the management of the user interface. In fact, the firmware also includes an automated calibration routine, to be triggered on demand. When invoked, this state machine sweeps the piezo offset voltage across the full  range to determine $\max R_1$ and $\max R_1$, as discussed about Eq.~\ref{eq:phitilde}. 

In general, user interaction is handled via a serial command parser, allowing real-time tuning of the sinewave generator parameters, PGA gaind, PID parameters, and monitoring the system status. While serial communication events have in general unknown durations, the architecture based on the global high-fidelity internal clock of the STM32H747XI microcontroller ensures that the 100 kHz acquisition and PID calculation are locked to a fixed time grid. In particular, if the serial communication events take longer than a T1 time cycle interval, the subsequent signal generation and acquisition operations are executed at the correct point in the time grid, without introducing spurious phase shifts in the harmonic signals.

\begin{figure}
    \centering
    \includegraphics[width=0.5\linewidth]{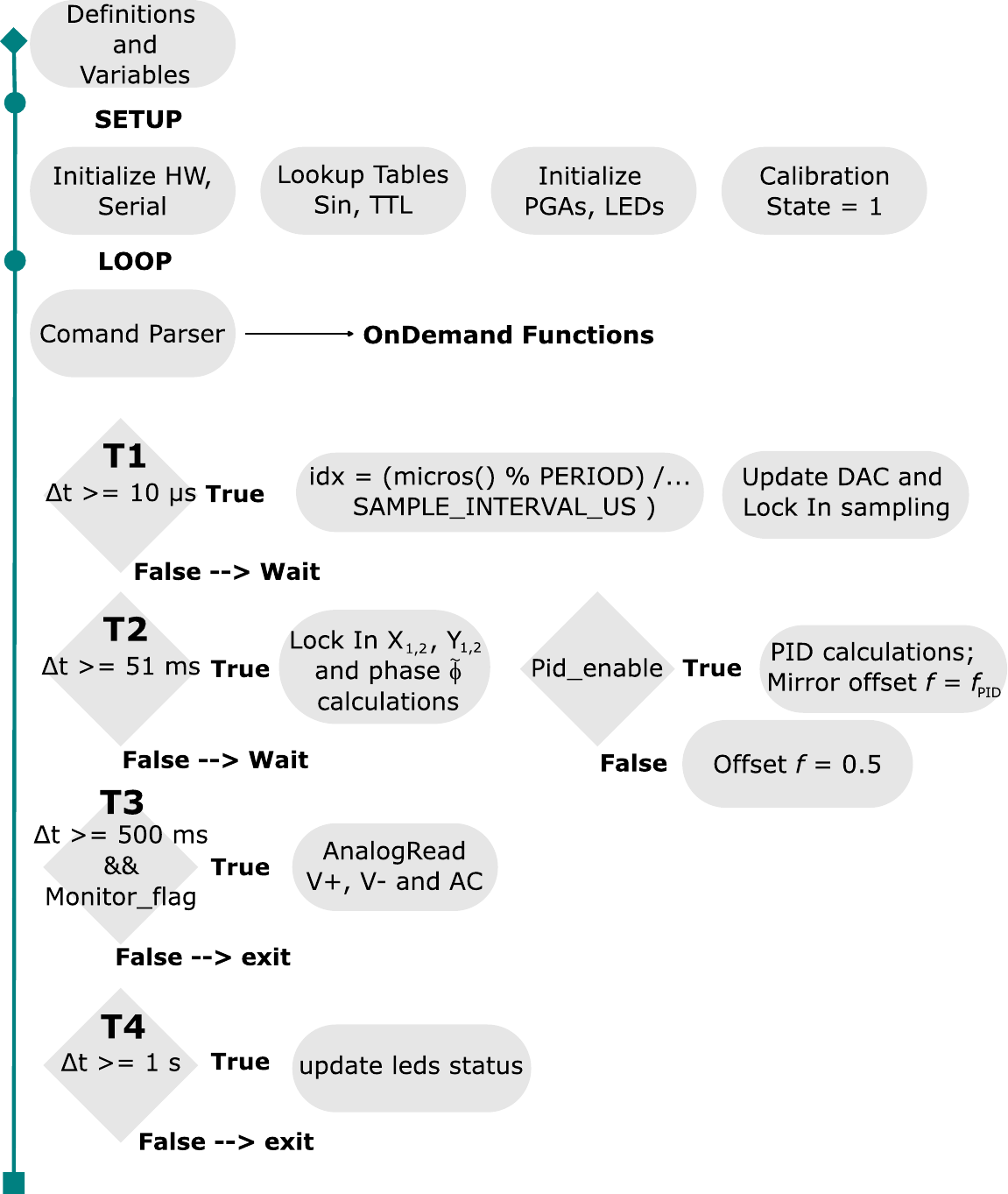}
    \caption{Flowchart of the firmware control algorithm. The system operates on a time-sliced super-loop architecture to ensure deterministic timing. (Top) The Setup phase initializes hardware peripherals, generates sine/cosine lookup tables for fast execution, and configures the PGAs via SPI. (Bottom) The Main Loop manages task scheduling via polling: T1 (10 $\mu$s) handles the critical ``Fast Loop'' for signal generation (DAC) and acquisition (ADC/Lock-in); T2 (51 ms) triggers the ``Control Loop'' for phase calculation and PID updates; T3 and T4 handle lower-priority monitoring and user interface tasks.}
    \label{fig3}
\end{figure}

\subsection{Python GUI}
To provide high-level control over the stabilization hardware, we developed a comprehensive Graphical User Interface (GUI) written in Python. The software is built upon the PyQt5 framework for the application layer and utilizes Matplotlib for real-time data visualization. Communication with the Arduino Giga R1 is established via a high-speed serial protocol (USB-CDC). The GUI operates on a multi-threaded architecture: a dedicated SerialReader thread handles asynchronous data ingestion to prevent freezing the interface during high-throughput transmission, while the main thread manages user interaction and plot rendering.
The interface offers several critical functions for experimental management: i) a real-time telemetry, where it continuously plots the in-phase ($X_1$,$X_2$) and quadrature ($Y_1$,$Y_2$) lock-in components, as well as the calculated phase $\tilde{\phi}$, allowing the operator to visually assess the lock quality; ii) a live tuning, where it allows controlling the input parameters and on-the-fly adjustment without resetting the microcontroller; iii) automated characterization, as the software includes a sequencer mode capable of toggling the feedback loop (PID ON/OFF) at set intervals. This feature, combined with an integrated data logging module that exports datasets to compressed NumPy (.npz) or CSV formats, streamlines the acquisition of long-term stability metrics and step-response tests.
Furthermore, the GUI allows: i) to directly analyze the collected data and to display them in a different window; ii) to open a secondary window (clicking on ``Monitor V+ V-'') that reports like a real-time oscilloscope the measured values referred to the DC and AC signal of the Monitor+ and Monitor-. Further details are given in Appendix \ref{app:ondemand}.
A critical design feature of the proposed system is the complete decoupling of the real-time control logic from the user interface. The Python GUI functions strictly as a supervisory telemetry tool; it visualizes data and updates parameter registers (e.g., PID gains) asynchronously. The stabilization algorithm is executed entirely on the Arduino Giga R1 M7 core; consequently, the feedback loop operates autonomously. Once the parameters are set, the USB connection can be disconnected or the host computer can be rebooted without interrupting the laser locking process. This headless operation capability ensures that the optical stabilization remains robust against operating system latency, software crashes, or USB communication timeouts. 

In addition to the Python-based GUI, we converted it in a standalone HTML web interface that eliminates the need for local software installation or Integrated Development Environment (IDE) configurations. This web application communicates directly with the interferometer stabilizer via a standard USB connection, offering a lightweight, cross-platform alternative. An example of the web interface is reported in Appendix \ref{app:webapp}.
\begin{figure*}[h]
    \centering
    \includegraphics[width=1\linewidth]{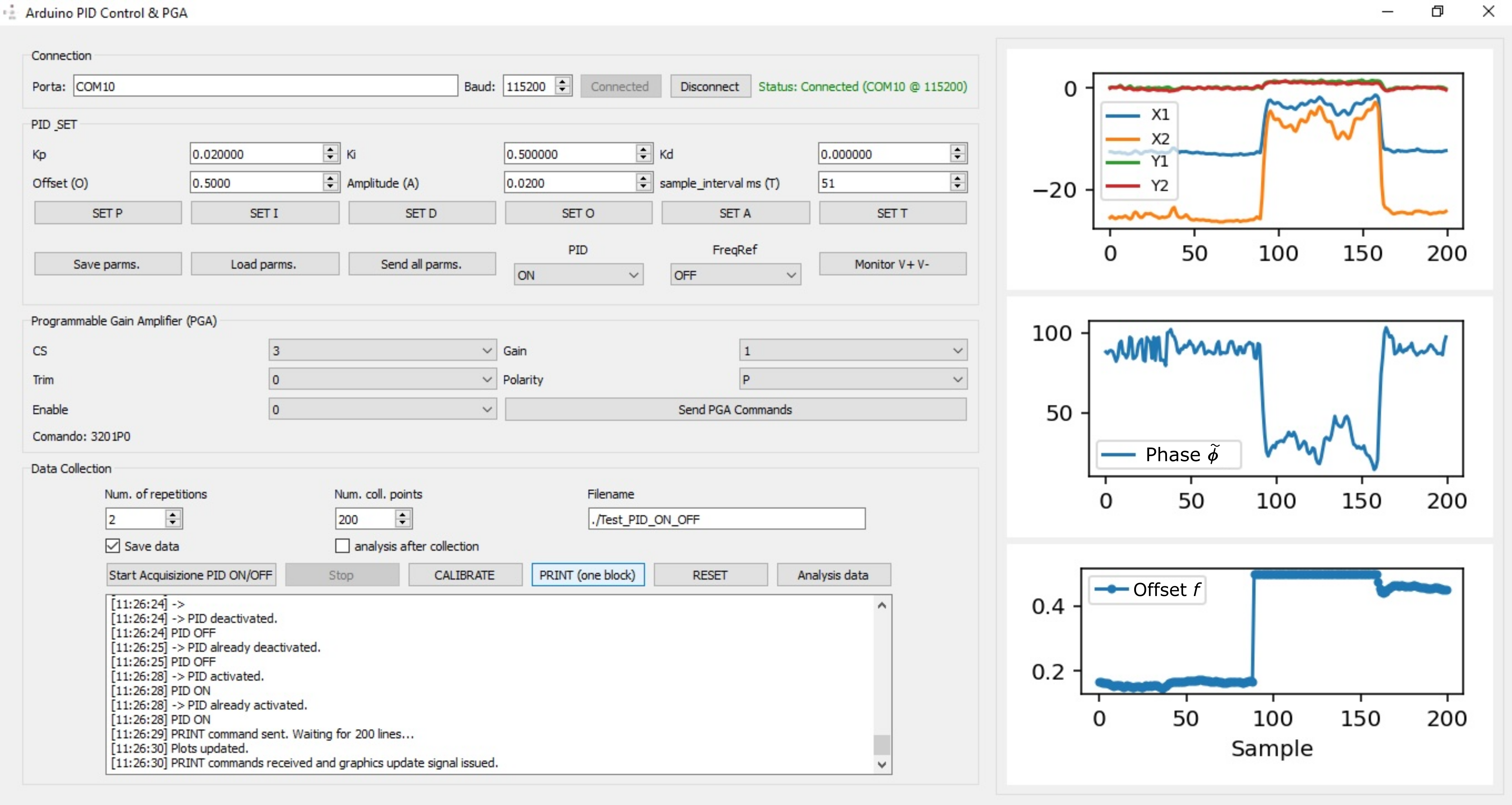}
    \caption{A screenshot of the custom-developed Python Graphical User Interface (GUI), which acts as a high-level supervisor for the Arduino controller. It provides real-time visualization of the interferometric signals (X,Y) and the estimated phase ($\tilde{\phi}$) via embedded Matplotlib charts. The control panel (right) allows for on-the-fly tuning of PID gains, modulation parameters, and the execution of automated stability sequences (PID ON/OFF) for performance characterization. The screenshot shows the measurements in different states with PID switched ON (samples 0-90), OFF (samples 90-160), and ON (samples 160-200).}
    \label{fig4}
\end{figure*}
\newpage
\section{\label{sec:discussion}Results and Discussion}

\subsection{Testing the calibration procedure}
\begin{figure*}[h]
    \centering
    \includegraphics[width=1\linewidth]{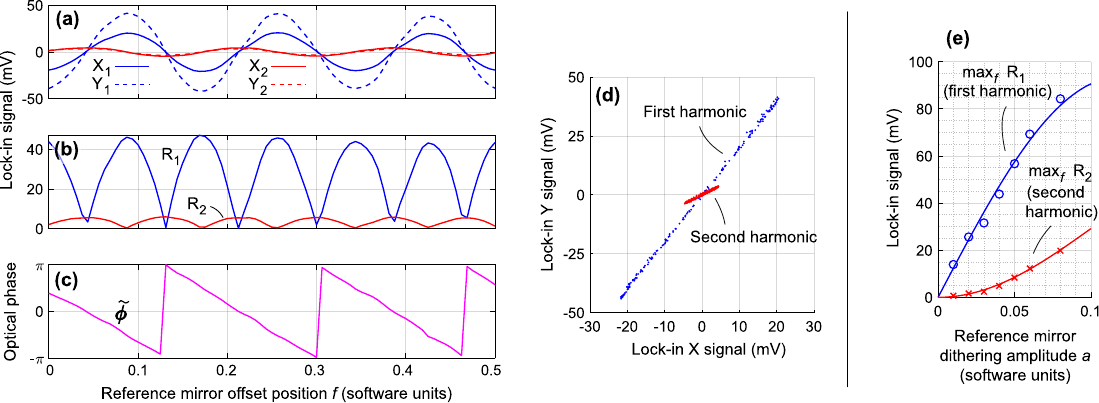}
    \caption{Testing the optical phase estimation protocol. Panels (a) and (b) illustrate the lock-in variables collected while the reference mirror is linearly swept, by acting on the control variable $f$. Sinusoids and absolute value of sinusoids are observed, according to the theory. Panel (c) reports the estimated optical phase, which behaves, as expected, as a piecewise linear function of $f$. Panel (d) reports a scatter plot analysis of the lock-in variables: the data for first and second harmonic are distributed in separate segments, as predicted by Eq.~\ref{eq:XY}. Panel (e) reports the maximum (with respect to $f$) of first- and second harmonic signals. Circles and crosses represent the measured values, while lines represent the theory (i.e., Bessel functions).}
    \label{fig5}
\end{figure*}

To demonstrate the proper operation of the phase detection system, we performed a first measurement where the reference mirror position $f$ is slowly and linearly swept in time (we imposed a sweep from 0 to 0.5 software units, i.e.~from zero to mid-scale, in 51 seconds ). Meanwhile, the reference mirror is dithered with a signal of amplitude $a = 0.04$, and the lock-in variables are collected. This has been done with the GUI by pressing the ``CALIBRATE'' button. The result is represented in Figure \ref{fig5}a-b, where it can be seen that the traces look like well-shaped sinusoids (or absolute values of sinusoids). This agrees with the theory, embedded in Eq.~\ref{eq:XY}-\ref{eq:R} in conjunction with the definition of the physical phase $\phi = 2(\Delta L - \delta_{r,0} )/\lambda$, with the assumption of constant $\Delta L$, and with the linear dependence $\delta_{r,0} \propto f$. Subsequently, from the $R_{1,2}$ traces (Figure \ref{fig5}b) we have determined $\max_f R_1$ and $\max_f R_2$, and by applying Eq.~\ref{eq:phitilde} we have determined the estimated optical phase, plotted in Figure \ref{fig5}c. As expected, the estimated optical phase $\tilde{\phi}$ depends linearly upon $f$.

Another way of representing the collected data and to test the compliance with the theory is proposed in Figure \ref{fig5}d. Here we have scatter-plotted $X_1$ versus $Y_1$ as blue points, and $X_2$ versus $Y_2$ as red points. According to Eq.~\ref{eq:XY}, one has that $Y_1$ is linearly dependent upon $X_1$, and similarly for the second-harmonic related quantities. This is indeed what has been observed, to a good level of accuracy. Moreover, from the representation of Figure \ref{fig5}d it is straightforward to determine the unknown, hardware-related phases $\psi_{\Omega}$ and $\psi_{2\Omega}$, since they are related to the slopes of the point clouds. However, such values are not relevant for the phase estimation protocol; for this reason we have not implemented their determination in the firmware. Instead, the firmware stores the $\max_f R_1$ and $\max_f R_2$, and updates them any time the calibration procedure is called.

Up to now, we have considered a fixed value of the dithering amplitude $a$. However, a significant impact of this parameter is expected, as the quantities $\max_f R_1$ and $\max_f R_2$ are, in theory, proportional to $J_{1,2}(2\delta_{r,M}/\lambda)$, where $\delta_{r,M} \propto a$. To check this behavior, we have performed a series of measurements at different levels of $a$, see Figure \ref{fig5}e. The data (points) are in excellent agreement with the Bessel function trend (lines), with a proportionality factor $\delta_{r,M} \approx 7.5\, \lambda\, a$. We also highlight that, in order to operate the ADCs in the center of their dynamic range, we have modified the gain settings of PGA2 and PGA3 as follows: when $a = 0.01$, $G_{\mathtt{PGA1}}=1$, $G_{\mathtt{PGA2}}=10$; in the other cases, $G_{\mathtt{PGA1}}=0.25$, $G_{\mathtt{PGA2}}=10$. The lock-in signals in Figure \ref{fig5} are obtained scaling the ADC counts with the PGA gains, hence yielding the physical voltage after the bandpass filter (see Figure \ref{fig2}). Nonetheless, such scaling is not needed to perform the phase estimate, as Eq.~\ref{eq:phitilde} is invariant upon scaling of $R_{1,2}$. Hence, to avoid unnecessary operations, the firmware elaborates all the lock-in variables in units of ADC counts, without scaling them to physical voltage levels.

\subsection{Testing the interferometer stabilizer}
The ultimate validation of the control system is its ability to maintain the interferometer at the linear operation point ($\tilde{\phi}\approx90\:^\circ$) against both slow environmental drifts and rapid external disturbances.

\begin{figure*}[t]
    \centering
    \includegraphics[width=1\linewidth]{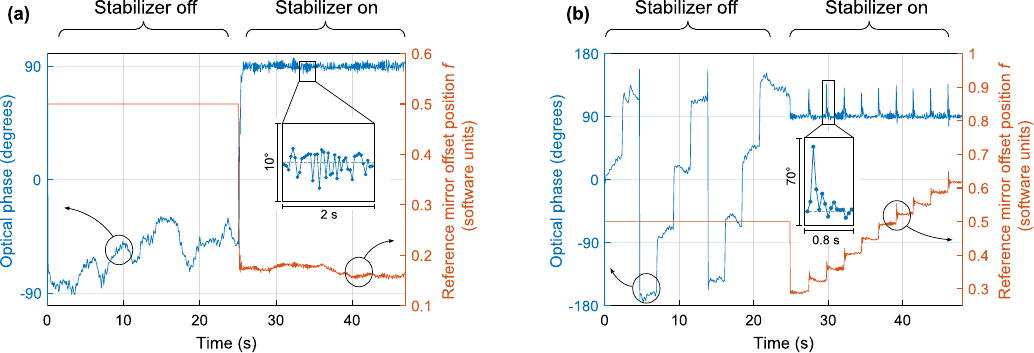}
    \caption{Closed-loop performance and disturbance rejection. (a) Stabilization against spontaneous (thermal) drift. The plot displays the temporal evolution of the interferometric optical phase $\tilde{\phi}$. In the first part of the graph (Stabilizer off), the system is free-running and exhibits a significant drift away from the working point ($\tilde{\phi}=90^\circ$). Upon activation of the feedback PID loop (Stabilizer on), the phase is rapidly locked and maintained at $\tilde{\phi}\approx90^\circ$ with a RMS error of $2.4^{\circ}$ . (b) Dynamic tracking capability under external perturbation. To simulate a step-like external perturbation, the reference mirror was mounted on a secondary piezo actuator driven every 2.5 seconds. The blue solid line shows the optical phase $\tilde{\phi}$, while the orange solid line shows the feedback actuation signal $f$. With the feedback disabled (left), the external motion induces large phase oscillations. When the PID feedback is engaged (right), the controller actively modulates $f$ to counter the disturbance, effectively clamping the phase to the set-point despite the continuous mirror motion.}
    \label{fig6}
\end{figure*}

Figure \ref{fig6}a demonstrates the suppression of long-term drift. The experiment begins in an open-loop state (Stabilizer off), where the PID feedback is deactivated, and the measured optical phase reflects the variations of $\Delta L$, which is subject to optical table thermal drift and to refractive index fluctuations induced by air currents. As observed in the first half of the trace, the phase $\tilde{\phi}$ drifts significantly, moving the system away from its linear sensitivity region. At approximately half of the experiment time, the stabilizer is activated. In a fraction of a second the controller adjusts the DC offset of the piezo mirror (i.e., the $f$ value), resulting is a sharp transition to a stable regime where the phase is locked to the target value. After the transition, the $f$ value oscillates, as it now compensates for the variations of $\Delta L$. Looking at the figure inset, the residual optical phase noise in the locked state can be assessed, and quantitatively estimated at $\Delta \tilde{\phi} = 2.4^{\circ}$ (RMS error). 

To further test the dynamic range and bandwidth of the feedback loop, we performed a forced-disturbance experiment, shown in Figure \ref{fig6}b. For this test, the previously fixed sample mirror was mounted on a secondary piezoelectric actuator driven in a stepped-like fashion (a quarter of a wavelength every 2.5 seconds). This setup mimics the effects, for instance, of a sudden mechanical failure, or of a change of sample position. Also in this case the experiments comprises two stages. In the first one the PID is deactivated (Stabilizer OFF), and as expected in the absence of feedback, the external mirror motion translates directly into optical phase excursions (Figure \ref{fig6}b, first data block). Then, enabling the PID (Stabilizer ON), the phase signal $\tilde{\phi}$ is instantly clamped to the target value. Crucially, the orange solid line of Figure \ref{fig6}b reveals the behavior of the actuation signal $f$: the controller generates a step-like counter-movement that compensates the external disturbance. Such compensation occurs on a time scale of the order of 0.1 seconds, as visible from the figure inset. This effectively cancels the optical path difference introduced by the moving mirror, proving the system capability to actively track and compensate for real-time mechanical perturbations. 

\section{Conclusions}
We have successfully built and tested a standalone, digital feedback controller capable of stabilizing a Michelson interferometer against thermal drift and mechanical perturbations. By leveraging upon the features of an Arduino Giga R1 board, by designing a dedicated analog signal board, and by implementing a coupled lock-in/PID firmware, we achieved an interferometric stability of a few degrees and a disturbance rejection bandwidth of approximately 10 Hertz. 

The primary advantage of the proposed approach lies in its accessibility. Commercial lock-in amplifiers and laser locking modules typically cost thousands of dollars, presenting a significant barrier for smaller laboratories or educational settings. On the other hand, custom FPGA-based custom architectures need elaborated programming techniques that are usually out of reach for optics researchers. In contrast, the system presented here — comprising the microcontroller and the custom interface shield — is advantageous both concerning the cost, as it can be assembled for few hundreds of dollars, and concerning the hardware/software complexity, thanks to well documented Arduino libraries and to our Github repository.

While the proposed system cannot compete on the performance metrics with intrinsically advanced electronic solutions, we highlight that in our proposal all the digital processing is performed on the Arduino board, as opposed to PC-based control loops, where system stability is tethered to the operating system performance and to the USB communication delays. Such a stand-alone device allows to lock the linear interferometric operating point even in the absence of a host computer, ensuring high reliability for long-term experiments or for harsh environments. 

Further improvements could also be likely deployed, such as the use of PCB designs including EMC compatibility, and advanced feedback protocols - possibly powered by neural networks. Nonetheless, we believe that the current implementation finds important applications in teaching workshops and in moderate-demand research laboratories.

\section*{ACKNOWLEDGMENTS}
We kindly acknowledge Francesco Francesconi (Università di Pisa) for precious advice about electronic filters. We acknowledge financial support by the Italian Ministry of University and Research (MUR) – PRIN Project ``TRUST'' – CUP B53D23004360006 – Grant Assignment Decree No. I53D23000570006.

\section*{AUTHOR DECLARATIONS}
\subsection*{Conflict of Interest}
The authors have no conflicts to disclose. 

\section*{DATA AVAILABILITY}
The data that support the findings of this study are available at the following link 
\url{https://github.com/GiuseppeELio/Arduino_Interferometer_Stabilizer}. Furthermore, the whole project is available on the Open Source Hardware Association (OSHWA) with the ID IT000026 under GPL-3.0 license. 


%

\appendix

\section{\label{app:circuits}Details of the analog filters}
\begin{figure*}[!h]
    \centering
    \includegraphics[width=0.7\linewidth]{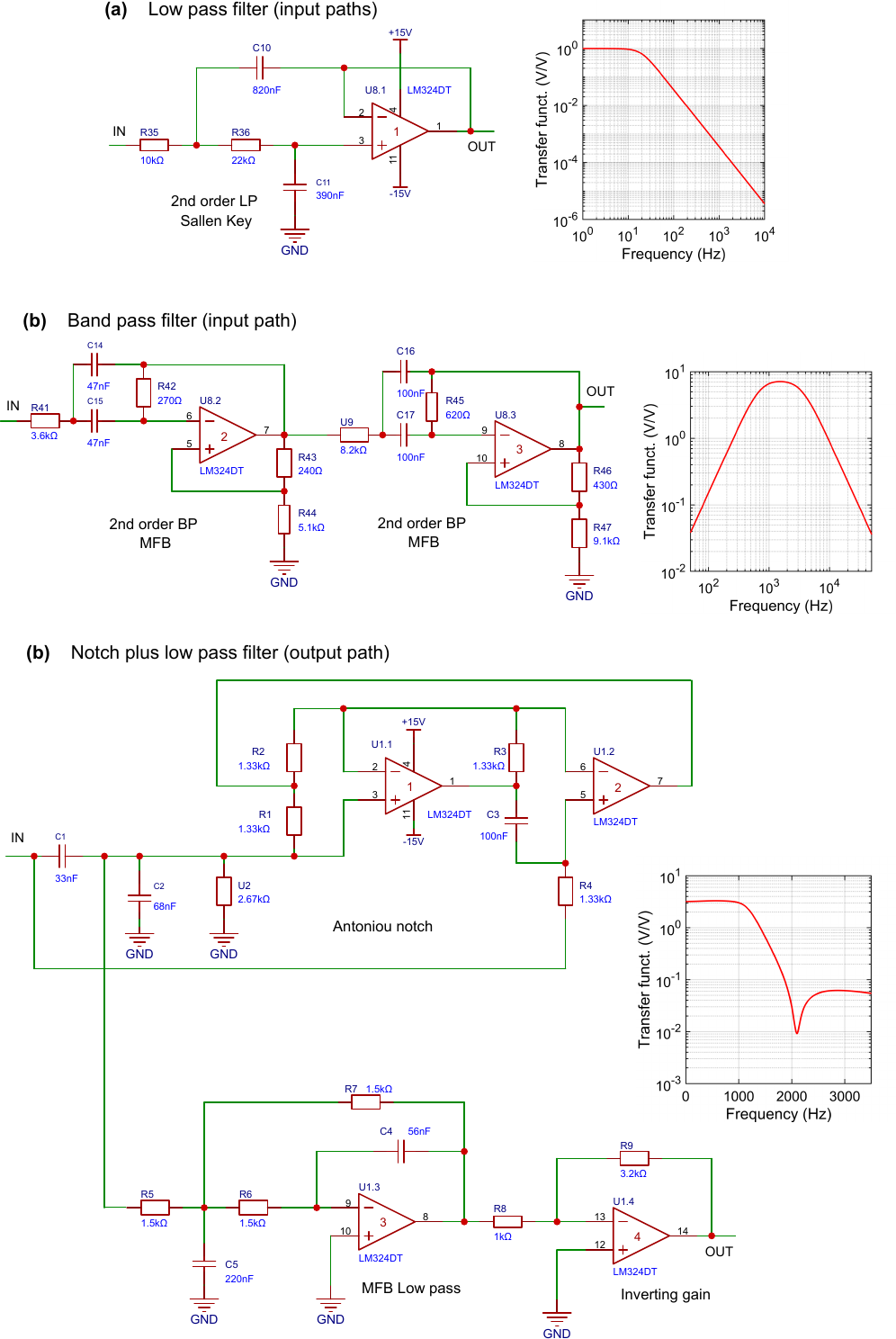}
    \caption{Schematics of the analog filters implemented on the stabilizer board. }
    \label{fig_CircuitSchem}
\end{figure*}

Figure \ref{fig_CircuitSchem} reports the schematics of the analog filters, alongside with the corresponding transfer function. The Sallen-Key and the Multiple Feedback (MFB) blocks have been designed with the Filter Design Wizard of Analog Devices\footnote{https://tools.analog.com/en/filterwizard}, and slightly adapted to the available components and manufacturing process. The notch filter based on the Antoniou synthetic impedance circuit is inspired by designs available in \cite{SedraSmith}.  
Here we have only reported the response function amplitude; a non-flat phase is also present. Such phase is in part responsible for a finite $\psi$ in Eqs.~\ref{eq:interf2} and \ref{eq:VA1}.

\section{\label{app:ondemand}Flowchart OnDemand Functions}

\begin{figure*}[!h]
    \centering
    \includegraphics[width=0.5\linewidth]{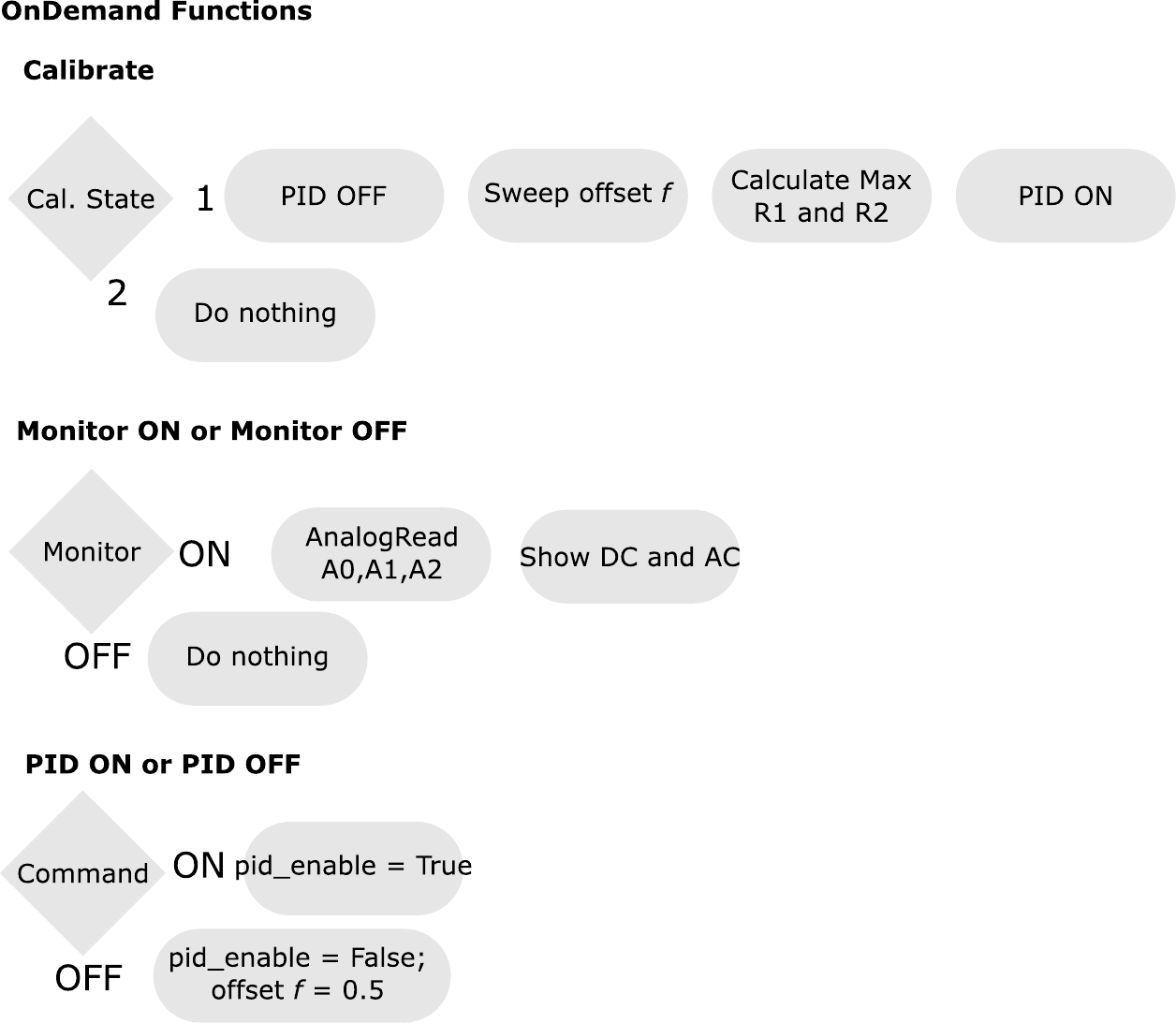}
    \caption{Flowchart of the secondary function recalled in the main firmware. The OnDemand functions are the ones that can be recalled by the command parsers directly by the terminal or by using the GUI interface. These functions are the calibration one, the toggle functions to activate/deactivate the monitor and/or the PID.}
    \label{fig_flow_ondemand}
\end{figure*}

Figure \ref{fig_flow_ondemand} reports a flowchart of the OnDemand functions. It is referred to the calibration state, the monitor ON and OFF function that allows to display or not a pop-up with information about the voltages in input for both the DC channels and the AC one. The flowchart reports also the toggle linked to the PID state that allows to activate or deactivate it when needed. 

\section{\label{app:calibration}GUI calibration procedure and Data Monitor}
\begin{figure*}[!h]
    \centering
    \includegraphics[width=1\linewidth]{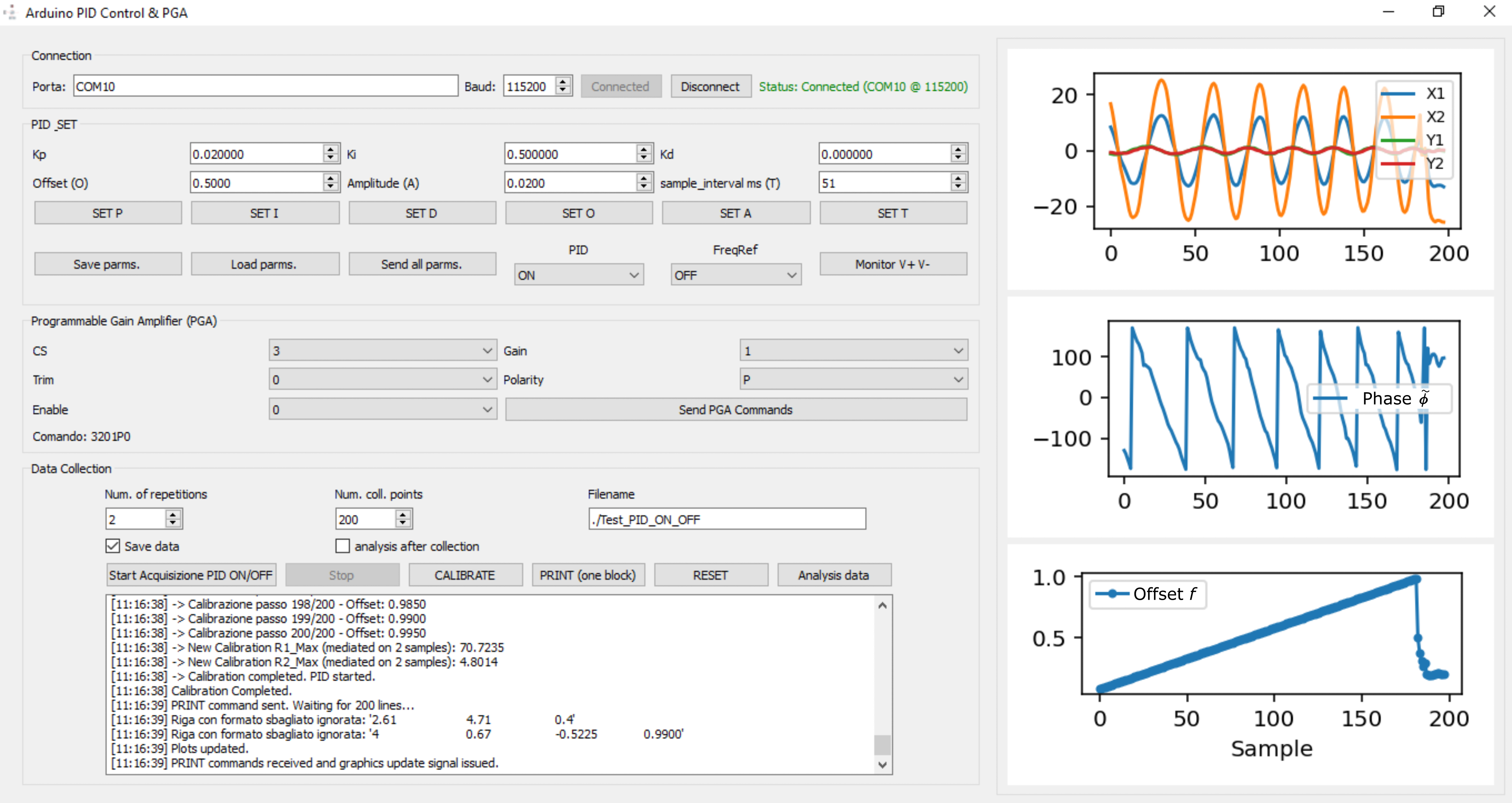}
    \caption{GUI interface showing the calibration procedure.}
    \label{fig_gui_cal}
\end{figure*}

This section reports other two examples of the GUI usage, the first one (see Fig.~\ref{fig_gui_cal}) is referred to the calibration procedure and how it appears on the user interface. The user can activate it just pressing the button CALIBRATE and it is automatically run over a 200 sample acquisition varying the voltage offset from 0 V to 1 V. The automatic analysis reports the values of X1, X2, Y1 and Y2 and the phase evaluation. 

\begin{figure*}[!h]
    \centering
    \includegraphics[width=1\linewidth]{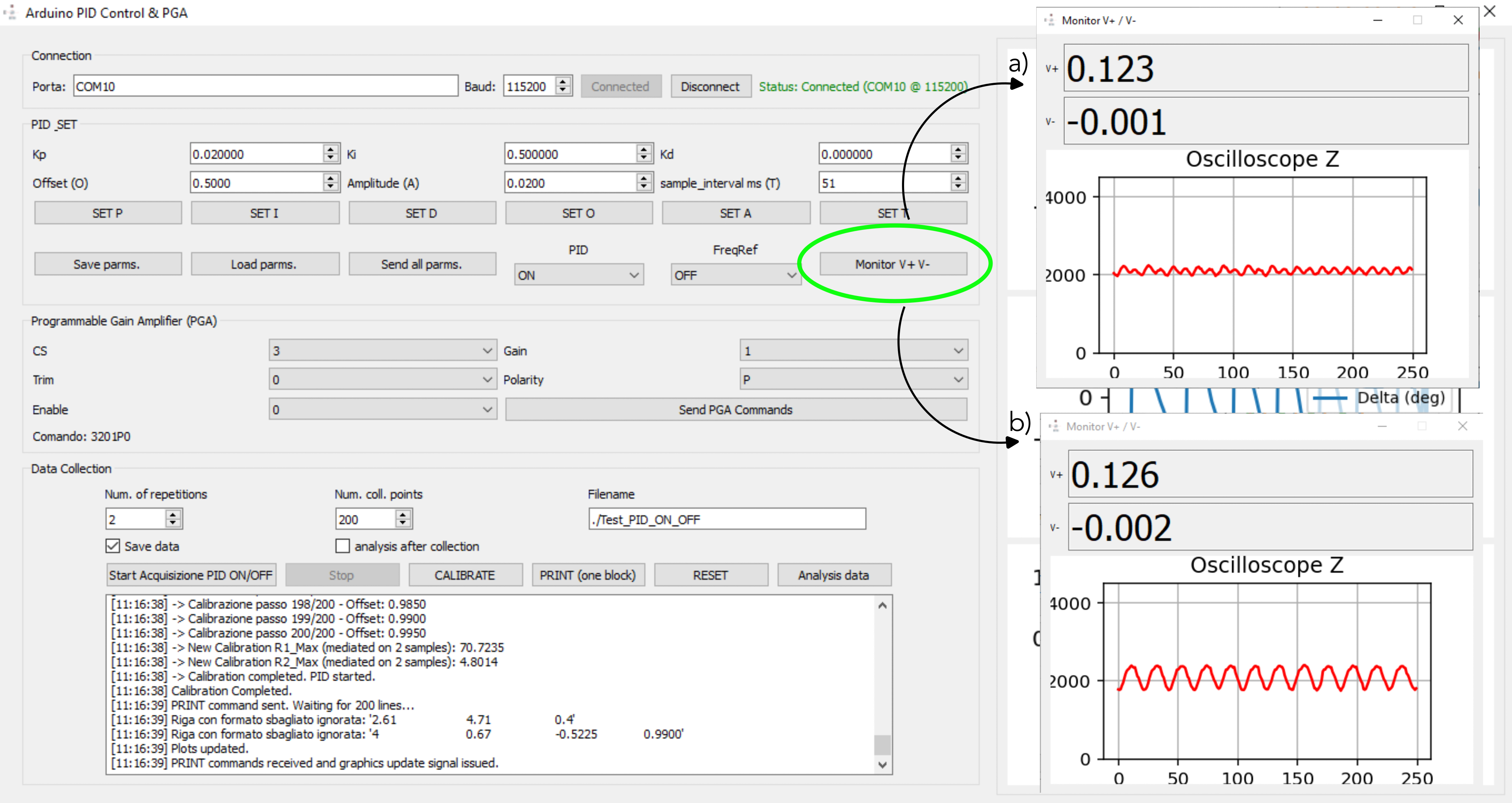}
    \caption{GUI interface showing the Monitor reporting the two DC channels (V+ and V-) and the live plot of the AC channel (an integrated digital oscilloscope). Two different conditions are illustrated: a) when the AC signal shows the presence of the second harmonic and b) when the AC signal shows the first harmonic.}
    \label{fig_gui_monitor}
\end{figure*}

The second one is referred to the monitor, see Fig.~\ref{fig_gui_monitor}. In this case the user can activate the monitor (the button in the green ellipse); then it is possible to immediately read the values of the two DC channels (V+ and V-) and to see a live plot of the AC channel, like an oscilloscope. The typical application is to diagnose if the signal displays mostly the second harmonic (\ref{fig_gui_monitor}a) or the first one (\ref{fig_gui_monitor}b). 

\newpage
\section{\label{app:characterization}Digital lock in amplifier characterization}
\begin{figure*}[!h]
    \centering
    \includegraphics[width=0.9\linewidth]{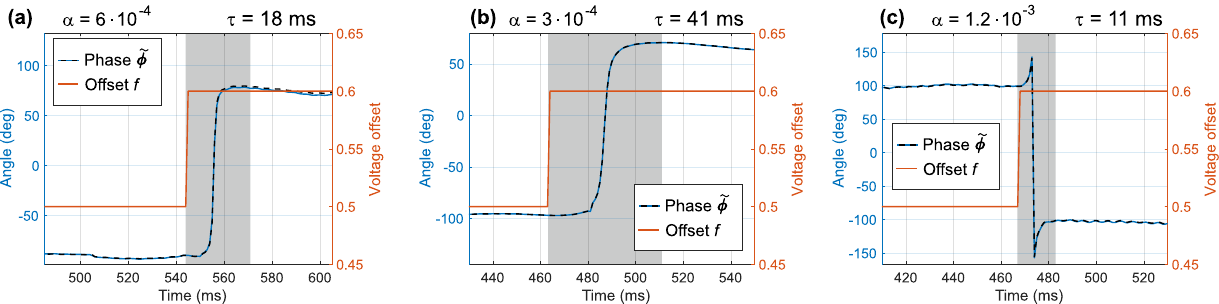}
    \caption{Experimental characterization of the digital lock-in amplifier dynamics. a) Signal evolution for the nominal filter coefficient $\alpha = 6\times 10^{-4}$, and a nominal response ($\tau\approx$ 18 ms), (b) Fast response ($\alpha/2$), and (c) Slow response ($2\alpha$). The varying rise times confirm the precise software control over the system integration bandwidth.}
    \label{fig_lockin_characterization}
\end{figure*}

In the main text, the lock-in amplifier stage is operated with a fixed digital filter parameter $\alpha = 6\times 10^{-4}$, which corresponds to a lock-in time constant $\tau \approx 18\, \mathrm{ms}$. This is in fair agreement with the theoretical prediction\cite{harvie23}, which states that $\alpha = \cos \gamma - 1 + \sqrt{\cos ^2 \gamma - 4 \cos \gamma +3}$, where $\gamma = 2 \pi /(\tau f_{s})$, being $f_s$ the sampling frequency. Such relation also states that an increase in $\alpha$ leads to a decrease in $\tau$, and vice versa. To validate this behavior, we tested the lock-in response to a step change of the mirror position, keeping the PID inactive. Figure \ref{fig_lockin_characterization} illustrates the time-dependent $\tilde{\phi}$ using different values of the filter coefficient $\alpha$. 
 Figure \ref{fig_lockin_characterization}a shows the response for the nominal case ($\alpha = 6\times 10^{-4}$), yielding a characteristic time constant $\tau\simeq$18 ms. We then repeated the test with halved and doubled integration times. As expected,  halving the coefficient (Figure \ref{fig_lockin_characterization}b) smoothed the response ($\tau\simeq$ 41 ms), providing higher noise rejection at the cost of temporal resolution, while, doubling the filter coefficient (Figure \ref{fig_lockin_characterization}c) reduced the characteristic time to $\tau\simeq$ 11 ms. This tunability allows the user to optimize the loop dynamics for specific environmental noise conditions without hardware modifications.

\newpage
\section{\label{app:webapp}Web Application}
\begin{figure*}[!h]
    \centering
    \includegraphics[width=0.7\linewidth]{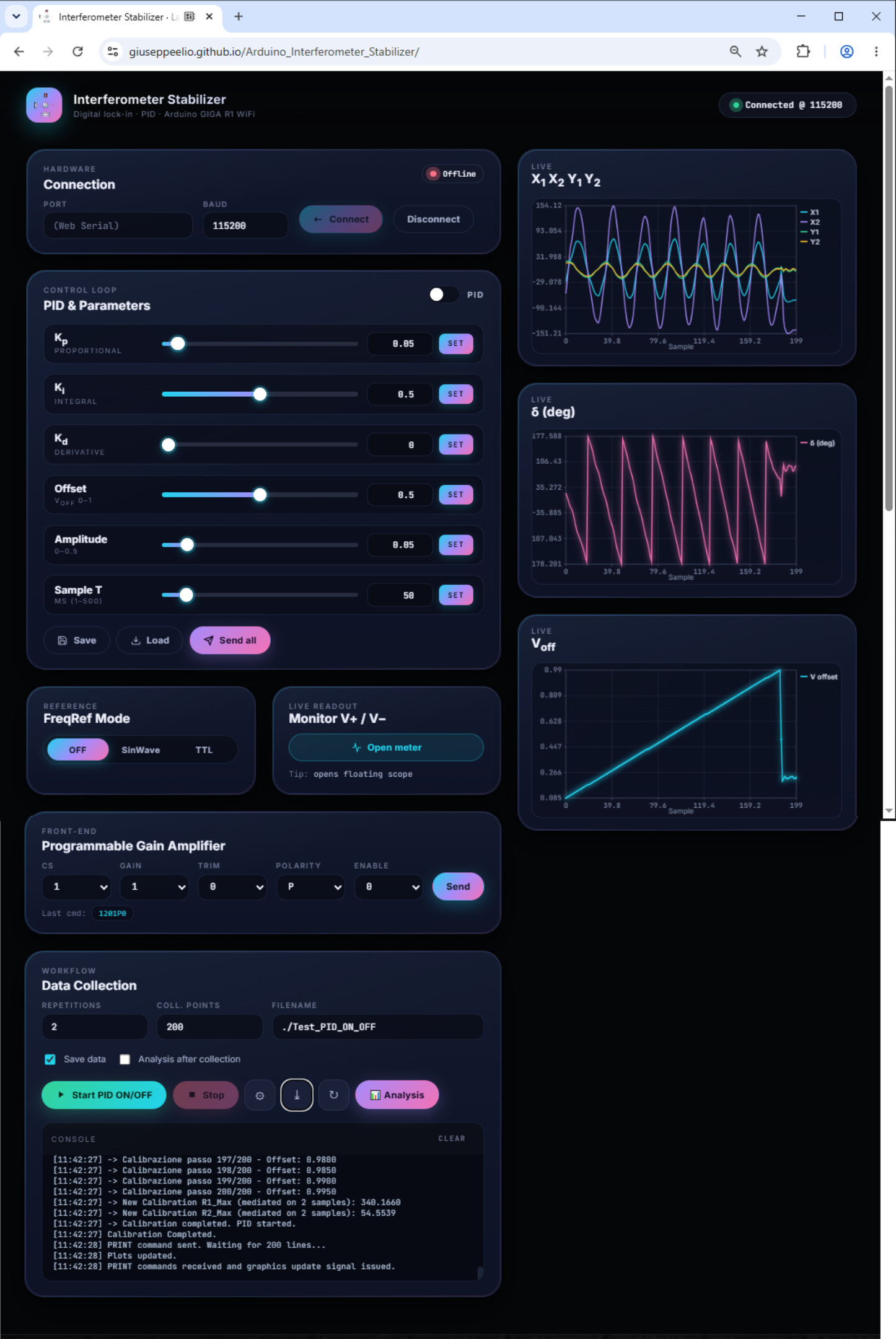}
    \caption{Web application directly hosted on GitHub web page. The example shows the calibration process.}
    \label{fig_webapp}
\end{figure*}

The web application has been translated from the Python GUI in order to work without any installed software just by using a web browser (Fig.~\ref{fig_webapp}). It has been translated from Python to HTML and Java by using Claude AI. The web application has been tested in a few different operational conditions resulting in stable operation. 

\end{document}